\def\year{2020}\relax
\documentclass[letterpaper]{article} 
\usepackage{aaai20}  
\usepackage{times}  
\usepackage{helvet} 
\usepackage{courier}  
\usepackage[hyphens]{url}  
\usepackage{graphicx} 
\def\UrlFont{\rm}  
\usepackage{graphicx}  
\usepackage[disable]{todonotes}
\usepackage{comment}
\usepackage{tabularx}
\usepackage{booktabs}
\newcommand{\citet}[1]{\citeauthor{#1} \shortcite {#1}}
\newcommand{\citep }{\cite}

\usepackage{fontawesome}
\usepackage{enumitem}
\usepackage[font=small]{caption}

\title{Learning to Classify Morals and Conventions:\\ Artificial Intelligence in Terms of the Economics of Convention}

\author{David Solans,\\
{Universitat Pompeu Fabra}\\
david.solans@upf.edu,
\And
Christopher Tauchmann,\\
{Technische Universität Darmstadt}\\
tauchmann@cs.tu-darmstadt.de, 
\And
Aideen Farrell,\\
{Universitat Pompeu Fabra}\\
aideen.farrell@upf.edu
\And
Karolin Kappler,\\
{FernUniversität in Hagen}\\
karolin.kappler@fernuni-hagen.de
\\\AND
Hans-Hendrik Huber,\\
{FernUniversität in Hagen}\\
hans-hendrik.huber@fernuni-hagen.de
\And
Carlos Castillo,\\
{Universitat Pompeu Fabra}\\
chato@acm.com
\And
Kristian Kersting,\\
{Technische Universität Darmstadt}\\
kersting@cs.tu-darmstadt.de
}

\begin{document}

\maketitle
\begin{abstract}
Artificial Intelligence (AI) and its relation with societies has  become an increasingly interesting subject of study for the social sciences. Nevertheless, there is still an important lack of interdisciplinary and empirical research applying social theories to the field of AI. We here aim to shed light on the interactions between humans and autonomous systems and analyse the moral conventions, which underly these interactions and cause moments of conflict and cooperation. For this purpose we employ the Economics of Convention (EC), originally developed in the context of economic processes of production and management involving humans, objects and machines. 
We create a dataset from three relevant text sources and perform a qualitative exploration of its content. Then, we train a combination of Machine Learning (ML) classifiers on this dataset, which achieve an average classification accuracy of 83.7\%. A qualitative and quantitative evaluation of the predicted conventions reveals, inter alia, that the \textit{Industrial} and \textit{Inspired} conventions tend to co-exist in the AI domain. 
This is the first time, ML classifiers are used to study the EC in different AI-related text types. Our analysis of a larger dataset is especially beneficial for the social sciences.

\end{abstract}

\section{Introduction} \label{sec:Introduction}

The term Artificial Intelligence (AI) describes a broad concept related to the ability of machines to carry out tasks in a way that might be perceived as “smart". Machine Learning (ML) constitutes a subfield of AI 
that studies algorithms that improve automatically through experience and have been used
to infer meaning, generalise and learn patterns from data and thus 
discover “knowledge" that was not explicitly programmed by the creator. 

In recent years, the AI domain has experienced an impressive growth.\footnote{https://www.forbes.com/sites/louiscolumbus/2018/01/12/10-charts-that-will-change-your-perspective-on-artificial-intelligences-growth/} 
Whereas the majority of the research around the concept of AI is concentrated on how to build more precise, reliable and advanced models, the main objective of this paper is to analyse advancements in and discussions around AI from a social sciences' point of view. From this perspective we examine which ‘conventions' or moral orders are employed during the creation of these models based on dialogue and justifications between individual(s) and the collective. By studying the research on, the design of, the development of and the public opinion on AI related systems, we focus on the interim process reasoned to be a key contributor to the subsequent interactions between humans and machines. 
To this end, we employ the Economics of Convention (EC) -- a general social science theory -- which proposes a pragmatic and situative perspective to study coordination and conflicts, analysing the underlying justifications and conventions.
Through the theoretical lens of the EC, we analyse how distinct moral registers represented by conventions within the EC are reflected in this domain. Having a better understanding of the conventions guiding the perceptions and advancements in the field of AI is considered to be a necessary preliminary step to a) understand the conventions reflected by these autonomous systems in their interactions with societies thereafter and b) shed light on ongoing conflicts around transparency or human vs. AI.

The Economics of Convention (EC) provide the framework for this study which are described in detail in the first part of this paper. For the analysis of conventions, we create a real-world text dataset with subsets from three different text sources and examine the distribution of conventions in these subsets. We use an iterative training process based on active learning as proposed in \citet{al_12} to build a supervised ML model with one binary classifier per convention and show results for each convention. 
The dataset along with the code is released to the research community.\footnote{Link to GitHub repository removed for anonymization. It will be provided in the camera-ready version of this paper.} 

\subsection{Objectives}
This work employs the theoretical framework of the EC to study written dialogues and research abstracts in 1) AI software design and development, 2) AI research and 3) social discussions around AI.  
Either researchers describe their findings to different communities (GitHub, Semantic Scholar (S2)) or AI is discussed in a community (Reddit). We aim to reveal the conventions, which these different communities follow.
We assume that documents in open-source ML and AI software repositories, and the conversations within, reflect the conventions guiding decisions taken during the AI development phase. Research articles in the domain of ML and AI describe findings to the research community and as such should reflect the conventions followed by scientists working in the field of AI research and design. For discussions in online forums where individuals with varied levels of expertise on the topic of AI exchange information and discuss recent advancements on the field we assume a broader and more general use of conventions.

\section{Structure of the Paper}
This paper is structured as follows: 
First, we provide an overview of the 
related work in relevant areas closely related to the work in this paper. After that, the theoretical framework of our analysis is described. The next section provides an overview of the creation of the dataset and the different subsets before we outline the architecture to train the ML models. In the subsequent section, we describe the results of the analysis of the dataset as we evaluate the performance of our classifiers and analyze the use of conventions in the different subsets of our dataset. We summarize our work in a conclusion, provide an outlook to future work and discuss limitations of our approach.


\section{Related Work} \label{sec:related_work}
Let us start off by providing an overview of the state of the art on the EC field. 
Research efforts focused in the analysis of each of the data sources considered in this work are summarized.

\subsection{Economics of Convention (EC)}
Although there is a large body of literature on understanding the motivation of open source software developers, none of them examines the use of the EC. \citet{Hurni2015} apply the EC in order to explain inter-organizational relationships in the coordination process of platform-based multi-sourcing in the general context of software development.
Non-technical approaches such as \citet{Denis2007}, \citet{Gkeredakis2014} or \citet{Kozica2014} use the EC to explain the coordination of pluralism and contradictory strategies in organizations.
Replacing the term “Economics of Convention” with “motivation” leads to additional results in the domain of software development. 
Especially in open source software development, several studies focus on motivation \cite{Hertel2003,Roberts2006}. Accordingly, previous research identifies five primary categories of motifs \cite{Bosu2019}: 
\begin{itemize}
    \item \textbf{Intrinsic motivation}, i.e., fun or self-efficacy \cite{Ryan2000}.
    \item \textbf{External rewards}, i.e., monetary incentives or career opportunities \cite{Lakhani2003}.
    \item \textbf{Ideology}, i.e., altruism \cite{Stewart2006}.
    \item \textbf{Community recognition}, i.e., fame or reputation \cite{Okoli2007}.
    \item \textbf{Learning}, i.e., development of personal skills or knowledge \cite{vonKrogh2012}.
\end{itemize}
However, these categories only partially relate to the EC, as the EC shifts the research perspective; the above mentioned along with most previous works rely on agent-based approaches, which focus on the agents or actors, while the EC studies situations, in which agents, objects, technologies, etc. interact.

\subsection{Content analysis of open source projects}
GitHub has been widely studied 
as a source of information for software development projects. Most of the existing contributions based on the analysis of open source project content fall under the following four categories: \textbf{user analysis}, \textbf{programming language prevalence}, \textbf{project quality analysis} and \textbf{project evolution predictability}. 
Due to the vast amount of studies on open source project content, this review is limited to contributions which are closely related to the work described in this paper.


Besides technical approaches, previous work on the study of project content often applies mathematical and statistical modelling to understand behaviour \cite{chen2017replicating}. This approach is also sometimes combined with qualitative studies based on automated processes. 
\citet{Sharma_2017} combine automated topic extraction with manual validation to 
categorise GitHub repositories based on the content of README files. Furthermore, \citet{Hassan_2017} propose the use of both qualitative and quantitative approaches to automatically detect instructions for software development in project description files.


Apart from these efforts, \citet{Prana_2018} automatically structure the content of GitHub README files. In order to do so, they combine manual annotation with automated text classification approaches. 
\citet{zhang2019explorative} perform a qualitative analysis of 
software projects related to scientific articles in the field of AI in work which analyzes content specifically related to ML and/or AI in GitHub.
Although there are studies on the content of GitHub project description files, these studies have different objectives.
In contrast, our work proposes for the first time the categorisation of AI and ML related projects based on the content of the README file according to the 
EC paradigm.

\subsection{Content analysis of scientific articles}

Although there is indeed much work in quantitative analysis on scientific articles, this body of work is mainly focused around the extraction of various entity and relation types such as named entities \cite{augenstein-etal-2017-semeval}, co-references \cite{gupta-manning-2011-analyzing} and semantic roles \cite{he-etal-2018-jointly}. 
Accordingly, previous work analysing Semantic Scholar (S2) focuses on those types \cite{luan_multi-task_2018}. Although there is work on the identification of patterns within the research community, this work is concerned with structural analysis such as citations and gender and not with discourse patterns \cite{vogel_he_2012}.
In recent work on language modeling in scientific texts, \citet{Beltagy2019SciBERT} report state of the art results on several standard NLP tasks. However, 
such a model is generally not directly feasible for convention classification as this complex task requires in depth control of the iterative labeling and classification process.


\subsection{Content analysis of online discussions}

Online forums and discussion sites are widely used to study social interaction. 
Different research communities study a variety of aspects such as the evolution and predictability of interactions in general \cite{Glenski_2017} and popular posts in particular \cite{Cunha_2016}. \citet{Buntain_2014} study the evolution of user communities and social roles. \citet{Bergstrom_2011} and \citet{Haralabopoulos_2014} focus on the reliability and correctness of the information. 

\citet{manikonda2017tweeting} perform a sentiment analysis of public perception of AI for expert and non expert groups of users on  Twitter and \citet{Javaheri2019PublicVM} compare opinions of the public and media on robots and autonomous systems. \citet{fast2016longterm} study the evolution of media perception of AI, and \citet{Manikonda2017WhatsUW} study privacy concerns of users about intelligent assistants by performing a survey and analysing public reviews.
While \citet{Datta_Adar_2019} study inter-community conflicts and common patterns, they define the conflicts as anti-social behaviour and do not consider the EC theory or other types of conflicts.

All this work proves that online social sites are valuable sources of knowledge for the understanding of social behaviours and opinions. Along this line, our work enhances the understanding of society's perception of AI through the EC framework.
\section{Economics of Convention (EC)}
The main focus of our work lies at the intersection of the EC theory and the research, design, development and public opinions of AI-related systems. The EC, as a general social science theory developed by \citet{boltanski2006}, proposes consistent pragmatic and situative concepts for the sociological analysis of behavioral coordination. It relies on 
justifications observed during ordinary disputes. This framework of justification is conceived as a theoretical research lens to empirically study cooperation and conflicts. In conflict situations, human actors mobilise arguments to defend their perspective. Based on field surveys and Western political philosophy, Boltanski and Thévenot develop a taxonomy of 
various conventions, or registers, of the so called “common good” the actors mobilize. The common good -- or the benefit or interests of all -- directly refers to specific perceptions of justice and fairness \cite{boltanski2006,diaz2018}. Hence, (potential) conflicts arise when 
a view of the common good that is based on one principle of justification is criticised according to criteria which underlie another principle of justification.
This theoretical approach has been already used in many different fields, e.g. the production of consumer goods \cite{storper1997,boisard2003} and health \cite{daSilva2018,sharon2018,batifoulier2018}. It is found to be useful for gaining more insight into what is at stake in emerging conflicts. \citet{boltanski2006} identify six justification registers, each based on different philosophical foundations in Western liberal societies and conceptions of justice and what is fair: \textbf{Civic}, \textbf{Industrial}, \textbf{Market}, \textbf{Domestic}, \textbf{Inspired}, and \textbf{Renowned}. \citet{boltanski2005} and \citet{lafaye1993} expand it with two more registers: the \textbf{Project} and the \textbf{Green} register. \citet{sharon2018} introduce a further \textbf{Vitalist} register based on the ‘googlization of health research'.
\begin{table}[t]
 \centering\small
 \begin{tabularx}{\columnwidth}{lll}
 \toprule
 \textbf{Convention} & \textbf{Common good} & \textbf{Values} \\
 \midrule
Industrial & Increased efficiency & Functionality, expertise, \\ & & optimization \\ 
 
 Project & Innovation &  Activity, experimentation,\\
 & and the network &  connection \\
 
 Market & Economic growth  & Competition, consumer \\ 
 & &  choice,  profit \\
 
 Inspired & Inspiration & Spontaneity,  deliberation, \\
 & & emotion \\
 
 Civic & Collective will & Inclusivity,  solidarity,\\
 & & equality \\
 
 Domestic & Tradition & Hierarchy,  trust \\
 
 Green & Protection of &  Environmental activism \\
 & environment   \\
 
 Renown & Public opinion & Popularity,  fame \\ 
 \bottomrule
\end{tabularx}
\caption{Registers of worth in the Economics of Convention} 
\label{tab:registers_worth}
\end{table}

Table \ref{tab:registers_worth} provides an overview of each of these registers with their principles of justification. It shows that there is a plurality of possible conventions or registers. The EC defines a ‘convention’ or ‘register’ not merely as a habit or custom \cite{thevenot2001,boltanski2006}; the concept of conventions in the EC is more complex. Conventions and registers form interpretative frameworks which actors develop and manage to evaluate and coordinate ‘action situations' \cite{diaz2019}. However, this does not imply that each individual is part of a particular convention, or that individuals consciously act according to the precepts of any of these mentioned \cite{daSilva2018}. On the contrary, depending on interactions with others, actors can easily pass ‘from one convention to another’ \cite{daSilva2018}. Similarly, the justifications for each of the actor's activities are implicit; individuals only make them explicit in a conflict. Coordination of these conflicts requires either agreement on a common principle or that the actors find a common understanding, which can then emerge between different registers of justification. All conventions refer to a legitimate and immeasurable conception of the collective so that no convention is more rational than any other. The decision for a certain convention or register is not merely a matter of calculation but a choice between several possible common traits the actors share in their interactions \cite{diaz2018}. Each register or convention acts as a logical, harmonious order of statements, objects and people that provide a general sense of justice. Hence, the typology of \citet{boltanski2006} offers an applicable framework to identify the conventions, which guide researchers, developers and their moral orientations in the field of AI.

\section{The EC Dataset}

\begin{table}
\centering\small
\begin{tabularx}{\columnwidth}{ll}
\toprule
 \textbf{Convention} & \textbf{Top keywords}  \\
\midrule
 Industrial & Performance, standard, tests, learning, reliable\\
 Project & City, projective, connections, links, networks
 \\
 Market & Customized, goods, license, sell, billion \\
Inspired & Inspiration, inspired, visual, passion, method  \\
Renown & Opinion, press, fame, audience, influence \\
Civic & Collective, civic, interests, license, children \\

Domestic & Superiors, upbringing, trust, dependence, origin \\

Green & Green, economy, growth, carbon, sustainable \\
\bottomrule
\end{tabularx}

 
\caption{\label{table:TFIDF_training} A combination of the top five keywords in the dataset per convention established by manual 
analysis and TF-IDF frequency
}
\end{table}

The dataset contains subsets from three main data sources: Semantic Scholar (S2) research paper abstracts\footnote{\url{https://semanticscholar.org}}, GitHub README files\footnote{\url{https://github.com}} and Reddit forums\footnote{\url{https://reddit.com/}}. 

To pre-filter documents we use a combination of two sets of keywords: 
First, we use a keywords list manually created by domain experts, including one of the authors and based on the registers introduced in Table \ref{tab:registers_worth}.
Second, we perform keyword matching after a first iteration of labeling based on ‘Term Frequency-Inverse Document Frequency' (TF-IDF) \cite{TFIDF_2010} to extract keywords that are more common for each convention and not so common for the rest. Table \ref{table:TFIDF_training} shows the five most important (of more than 30) keywords for each convention. 


\subsection{GitHub}
GitHub 
is a web-based interface and cloud-based service that provides tools to effectively store and manage code in addition to tracking and controlling changes in the code base. GitHub stores the code and metadata of more than 100 million projects with involvement from more than 31 million developers.\footnote{\url{https://github.blog/2018-11-08-100m-repos}} 
More than 8,500 projects related to AI topics are collected using the official GitHub API. We collect the content of the README file along with creation and last update timestamps in addition to statistics about the popularity of a repository. To avoid bias, repositories from all different levels of popularity ( measured with the GitHub star rating) are gathered.
In order to compare the use of conventions in GitHub AI related repositories with those in non-AI related repositories, data from an equivalent number of repositories similar to AI related topics is collected. Similarity is calculated on the basis of the number of stars. 
Table \ref{tab:datasets} shows the no. of sentences and the no. of repositories in the GitHub subset.

\begin{table}
\centering\small
\begin{tabularx}{0.8\columnwidth}{lll}
\toprule
 \textbf{Data source} & \textbf{Sentences} & \textbf{Items}  \\
\midrule
 GitHub AI & 127,236  & 8,609 repositories\\
 GitHub non-AI & 71,706 & 5,358 repositories\\
 S2 AI  & 22,742 & 2,954 abstracts \\
 S2 non-AI & 69,694 & 5,970 abstracts\\
Reddit AI & 38,296 & 2,455 threads\\
Redit non-AI & 219,916 & 3,875 threads \\
 \midrule
 Total size & 549,590 & 29,221 \\
\bottomrule
\end{tabularx}
 
\caption{\label{tab:datasets} Counts of sentences and items for AI and non-AI subsets from each data source. Depending on the specific data source, items refer to repositories, abstracts or threads.} 

\end{table}

\subsection{Semantic Scholar (S2)}

Semantic Scholar (S2) is a search engine for peer-reviewed articles, which provides an open research corpus with more than 40 million papers from computer science and bio-medicine in machine readable JSON format \cite{ammar:18}. For the analysis of the conventions, we select a sample of entries that appear in one of the AI conferences listed in \cite{kersting2019ki} and which are published after the year 2016. This list helps us to analyze the use of conventions in different sub-fields of AI, such as robotics, computer vision and natural language processing. We only select publications from 2016 onward because during this time, research in AI and applications of ML in particular received a significant boost with the release of TensorFlow \cite{tf_16}. This sample is further narrowed down by pre-filtering documents with the help of a list of keywords that belong to either of the registers in Table \ref{tab:registers_worth}. Table \ref{table:TFIDF_training} shows some of the most important keywords from this list. 

\subsection{Reddit}

 Reddit is a website centered around social news, web content rating, and discussion. Communities are named ‘subreddits' and created around topics.
We collect different threads from ML and AI ‘subreddits'. In detail, the text from the title of post which starts a thread, its body and the first level answers are collected by using the Reddit API. Samples from the AI domain are collected from a ‘subreddit' called ‘r/artificial', whereas the non-AI examples were gathered from a variety of ‘subreddits' related to the  computer science field: ‘Javascript', ‘DataBase', ‘Python', ‘Android'. 
We only use threads with a minimum of 4 upvotes (positive votes by readers from the community) to ensure that only relevant threads are considered in the analysis.
 


\section{Methods for building the EC Model}
\label{Classification Model Architecture}
In order to build an EC ML model and analyse the predictions on our dataset, we define the EC classification as a multi-label task whereby each sentence in our dataset may have multiple associated conventions and hence multiple labels.


To the best of our knowledge, this is the first attempt to build a text-based EC classifier and no existing datasets can be used to train such an ML classifier. We regard the creation of a dataset for this purpose as a valuable contribution to the scientific community. Due to the complexity of the EC theory, the labeling of the dataset facilitated by the authors of this paper was a time consuming task necessitating expertise and care. To optimize the labeling effort we use an active learning approach \cite{al_12} focused on the labelling of items most beneficial to the training of the models. The quality of the predictions are thus incrementally improved while at the same time new samples are labeled to train successive versions of the classifiers.


\subsection{Model selection} \label{Model selection:}

The EC model should cover the following:


\begin{itemize}
\item Support multi-label classification, where one sentence can have multiple labels and the number of labels per sentence is not fixed.

\item Support multi-class classification, where sentences can 
belong to 1 out of multiple categories


\end{itemize}

To this end, the classifiers are trained using a strategy commonly known as one vs. rest (or one vs. all) \cite{Rifkin_2004}. This strategy involves the training of one binary classifier per class (i.e. convention) to model a multi-class problem.
As such, the eight binary class-labels show multiple classes per item (i.e. sentence) along with a confidence score between 0 and 1 for each predicted label. 
This in effect 
represents a multi-label architecture because one item can belong to multiple classes (i.e. one sentence can belong to more than one convention). We decompose a multi-label, multi-class problem into a set of binary classifiers.
The upside of the one vs. all strategy is that it enables classifier calibration in terms of precision. Selecting a classification threshold with equal levels of precision for all classifiers allows a balanced comparison of the results from the different classifiers. A classifier only outputs a positive label when this threshold is exceeded, otherwise the label is negative. 
%
Furthermore, the architecture based on classifiers that are combined into one big model facilitates the building and testing of individual convention classifiers which offers individual performance checks. This lightweight approach also eases the data handling process in the active learning scenario.
We use 
convolutional neural network (CNN) classifiers  
following 
 the architecture proposed by \citet{Kim_2014} with the standard parameters. 
The network uses an input sequence of 32 vectors per sample to represent a sentence, where each of the vectors is encoded with a 100-dimensional word embedding vector. The network is composed of 14 layers, four of them convolutional layers, with 
over 10 mio. parameters 
of which $\sim$ 300k are trainable.
It uses \textit{categorical cross entropy} as loss 
and a \textit{relu} activation function for the hidden layers. 

Accordingly, one individual classifier $C_c$ is trained per convention $C$. Given a sentence $S$, the classifier $C_c$ is trained such that it assigns a probability score $P$ for that sentence being part of the convention $C$. Therefore: $C_c (S,C) = P$ where $P=[0,1]$.
A combination of $N=8$ binary classifiers (one per convention) 
predicts the probability of an item (sentence) to belong to each possible class label (convention). We set the calibration threshold to 0.9 precision during training to ensure meaningful labels.
We classify conventions on sentence level because 
sentences correspond to the minimal units which reflect conventions in text.

As a ML classifier requires data input in the form of numeric values rather than continuous or discrete variables, a method to numerically represent the training text in the form of a vector is required. The most common approach to date to solve this problem is the use of word embeddings. Words are transformed into n-dimensional vector representations and projected into a new multidimensional space. The contextual relationship of words with similar context is reflected in the n-dimensional space by distance (e.g. similar words are close to one another). 
To this end we use pre-trained GloVe word embeddings \cite{pennington-etal-2014-glove} for the vector representation of words in this n-dimensional space.

\begin{figure}
  \centering
      \includegraphics[trim=0 30 0 0, width=\columnwidth
      ]{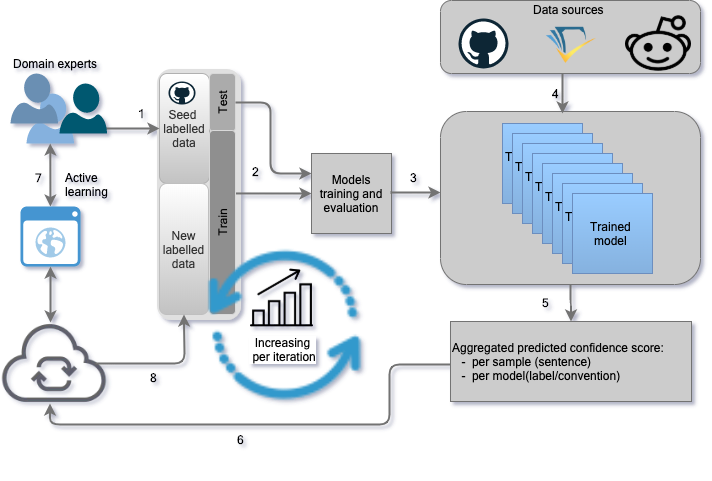}
  \caption{
  Active learning pipeline to 
  collect 
  and 
  verify 
  training data}
  \label{fig:data_bootstraping}
\end{figure}

\subsection{Labeling of dataset and active learning}

Due to the complexity of the EC, labeling the dataset demands both time and expertise. That is why an active learning model with a focus on uncertainty sampling is implemented. Uncertainty sampling prioritizes correctly labelling items based on classifier confidence. One objective is to enhance the training data by correctly labelling items that are classified with a low confidence score below 0.2 and improve classifier performance like that. Further focus is on correctly labeling items classified with a confidence close to the classifier's decision boundary (i.e. between 0.4 and 0.6) and a strong focus lies on confirming the models' belief in items with a confidence score above 0.8).
A total of 60\% of the labeled samples in our dataset come from high confidence predictions, 35\% are (re-)labeled from the low confidence predictions and the remaining 5\% 
come from the interval around the decision threshold. 

The models are updated with an iterative active learning pipeline 
After each iteration the model is evaluated on a fixed labeled set of items of 20\% of the (growing) entire dataset. A fixed set is suitable for fast evaluation. The pipeline illustrated in figure  \ref{fig:data_bootstraping} includes the following steps:

\begin{enumerate}[label={(\arabic*)}, noitemsep, leftmargin=0.6cm
]
\item The classifiers are pre-trained with seed data. To this end, domain experts labeled a random set of sentences from the GitHub subset.
\item In the first iteration, the eight classifiers are trained with the seed data, new labels are incorporated in suceeding iterations.
\item The performance of the trained classifiers is evaluated on labeled data and 
they are ready for predictions on unseen data.
\item Sentences from GitHub, S2 and Reddit are classified.
\item The 
classification outputs eight confidence scores per sentence (one per classifier). 
\item The aggregated data containing sentences and the associated confidence scores is pushed to a centralised cloud service and consumed by our web based active learning tool\footnote{A Python-based interactive GUI}.
Since the labeled data should be representative 
of the available unlabeled data, 
The active learning tool 
shows a histogram 
to provide insight to the most beneficial areas of focus for the domain experts.
\item Domain experts validate or relabel sentences with a confidence score or label unseen sentences.
%
\item The labeled sentence is 
added to the training data for the next iteration. 
A separate algorithm 
ensures 
equal numbers of positive and negative examples per classifier to avoid imbalance.
Steps (2) to (8) are repeated until training data suffices.
\end{enumerate}


\begin{table}[t!]
    \centering\small
     \begin{tabularx}{0.9\columnwidth}{ lllll}
     \toprule
     \textbf{Convention} & \textbf{Accuracy} & \textbf{AUC}  & \textbf{N}  & \textbf{$E_{prevalence}$}  \\
     \midrule
    Industrial & 0.750 & 0.708 & 1289 & 1/10 \\

    Project & 0.801 & 0.828 & 521 & 1/100 \\

    Market & 0.870 & 0.931 & 1082 & 1/100 \\
    
    Renown & 0.812 & 0.859 & 301 & 1/100 \\
    
    Civic & 0.902 & 0.897 & 477 & 1/1000 \\

    Inspired & 0.801 & 0.895 & 355 & 1/1000 \\
    
    Domestic & 0.866 & 0.901 & 475 & 1/1000 \\
    
    Green & 0.901 & 0.931 & 280 & 1/10000 \\

    \bottomrule
    \end{tabularx}
    \caption{\label{tab:Models_quality} Comparison of model performance per convention}
    
\end{table}
\begin{table}[t!]
    \centering\small
     \begin{tabularx}{0.6\columnwidth}{lll}
     \toprule
     \textbf{Data source} & \textbf{Accuracy} & \textbf{AUC} \\
     \midrule
    GitHub & 0.792 & 0.823 \\

    S2 & 0.748 & 0.749\\

    Reddit & 0.789 & 0.765\\
    
    \bottomrule
    \end{tabularx}
    \caption{\label{tab:Models_quality_datasource} Model performance per data source}
\end{table}

We ensure label quality with quality checks using a Qualitative Data Analysis (QDA) software\footnote{https://atlasti.com/} following the principle of deductive procedure for content analysis \cite{Mayring2014} parallel to the iterative active learning pipeline approach. We ensure the validity and reliability of the qualitative analysis by means of investigator triangulation. Investigator triangulation involves the use of multiple researchers in an empirical study \cite{Archibald_2016}. 
Our investigator triangulation involves 
three authors of this paper from different disciplines in the coding and labelling process and external EC-experts, with whom codes and labels are contrasted and discussed. The final coding iteration is performed on a random 
sample of 100 threads per data set, including context information such as links to the original posts in order to account for the situational approach of the EC.

\begin{figure}
  \centering\small
      \includegraphics[width=\columnwidth]{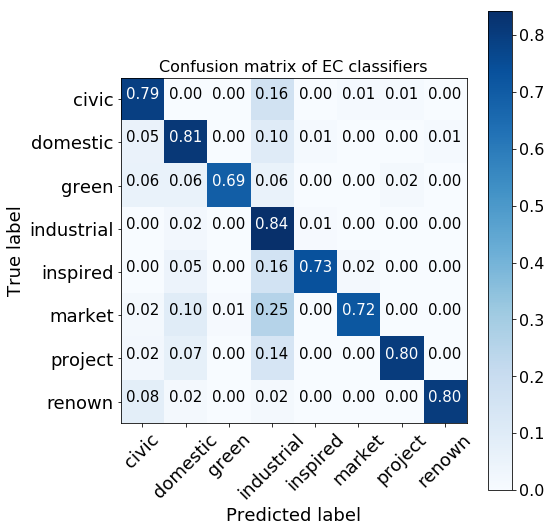}
  \caption{Confusion matrix of EC classifiers}
  \label{fig:EC_Conf_matrix} 
\end{figure}

\section{Results} \label{sec:results}
This section evaluates the performance of the classifiers on the entire dataset as well as on each subset. Furthermore, we present a quantitative and qualitative analysis of the predicted conventions.

\begin{figure*}[t!]
      \includegraphics[trim=0 4 0 0,width=\textwidth]{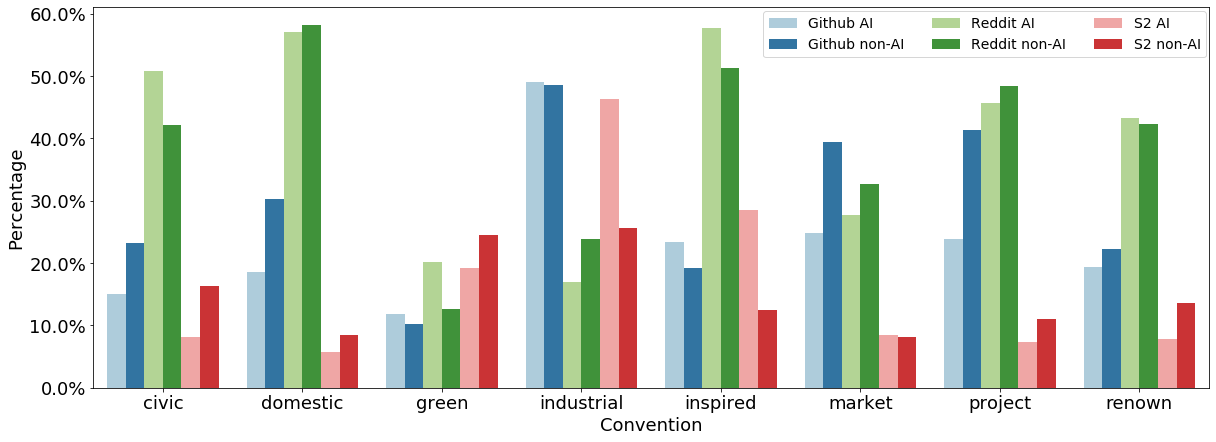}
\caption{Percentage of conventions in each data subset for AI and non-AI related items as predicted by the classifiers.}  \label{fig:Github_convention_proportions}
\end{figure*}
\subsection{Performance of classifiers}

We evaluate the performance of the classifiers with the following metrics:\begin{itemize}

\item{\textbf{Accuracy}: Accuracy is the ratio of correctly predicted elements between all the samples. Accuracy measures the ability of the classifier to identify elements from the positive and the negative classes and also considers the ability to differentiate positive samples from the negative ones.}

\item{\textbf{Area under curve (AUC)}: The AUC score provides an aggregate measure of performance across all possible classification (confidence) thresholds. AUC can be interpreted as the probability for a model to rank a random positive example higher than a random negative example.}

\item{\textbf{Precision}: Precision is the ratio $tp / (tp + fp)$ where $tp$ is the number of true positives and $fp$ the number of false positives. Precision is intuitively the ability of the classifier not to label as positive a sample that is negative. Precision is used to set the performance acceptability threshold for the built classifiers.}

\end{itemize}

Each of the models is independently evaluated on the test set with both metrics using leave-one-out cross validation. For each classifier, a classification threshold with value $T_{calibration}$ is selected so that at least precision of 90\% in test is obtained. Having similar precision for all of them facilitates the comparison of their predictions and ensures a limited amount of false positives.

Table \ref{tab:Models_quality} contains the average score for each classifier according to the following metrics: the number $N$ of training samples for each convention and a value $E_{prevalence}$ referring to the estimated prevalence of each convention in the dataset, which we determine in a manual analysis. Only a small number of conventions with a high discrepancy between $N$ and $E_{prevalence}$ are in the dataset, so we collect samples from other data sources to train such classifiers. 
Learning curves provide insight about the amount of labeled data which the classification models require to achieve satisfactory results and the amount they need to improve the results. 
We use ten fold cross-validation to split the whole dataset $k=10$ times in training and test set. Accordingly, the classifier is trained repeatedly on all but one of the subsets 
and evaluated on each one of the other subsets and a score for each training subset size and the test set is computed. Afterwards, the scores are averaged over all k runs for each training subset size.


In order to show that the classifiers generalize across all data sources, we calculate their performance for each individual data source. Table 
\ref{tab:Models_quality_datasource} shows average scores on equal numbers of positive and negative examples per convention. 
We see very similar performance across data sources.

A confusion matrix illustrates how well each classifier differentiates between positive and negative samples. The diagonal represents the ratio of true positives whereas the rest of the matrix corresponds to false negatives. Rows of the confusion matrix are normalized by using the total number of examples having a certain true label, so numbers represent the percentage of samples from each convention matched by each classifier. 
Figure \ref{fig:EC_Conf_matrix} shows the confusion matrix for each classifier using the $T_{calibration}$  threshold. To create the confusion matrix we select only sentences with a single label. Values in the cells represent the amount of sentences matched by each classifier for each convention. High values between 0.6 and 0.92 accuracy are in the diagonal axis of the matrix -- the classifiers are correctly differentiating. The Classifiers for the \textit{Civic} and \textit{Market} conventions are performing best.

\subsection{Evaluation of conventions}

\begin{figure*}
    \centering\small
      \includegraphics[width=0.9\textwidth]{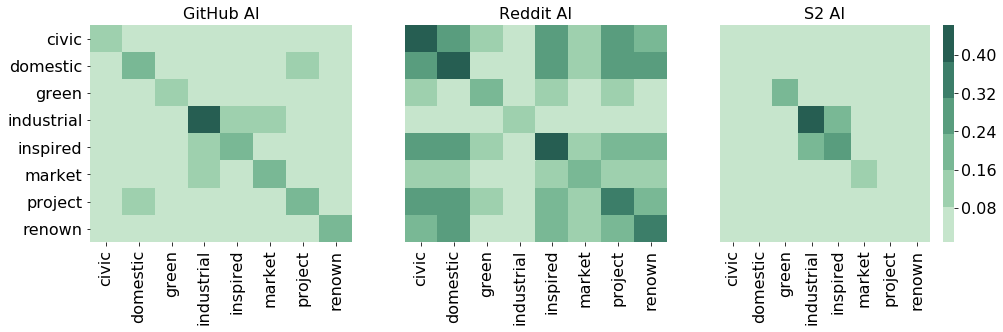}
\caption{Co-occurrences of conventions in the predictions for AI subsets. Values in the matrices are normalized by the number of sentences in each data source.} 

\label{fig:convention_cooccurences}
\end{figure*}




In the following evaluation, we discuss our EC classification results, compare the conventions in AI and non-AI subsets of our dataset, and present the co-occurrences of conventions.

Figure \ref{fig:Github_convention_proportions}
shows the distribution of both AI and non-AI related sentences for each data subset. In general, the prevalence of the different conventions is fairly aligned with the estimated ones. Regarding the different conventions, the \textit{Industrial} convention is very dominant in Github (AI and non-AI) and S2 (AI) with a proportion of about 50\%. As Github consists mainly of technical descriptions and standards and S2 of scientific abstracts, this is in line with our expectations. In S2, the \textit{Civic}, \textit{Domestic}, \textit{Market}, \textit{Project} and \textit{Renown} conventions are rarely present, while the \textit{Inspired} convention refers to innovative approaches and the \textit{Green} convention links with ecological projects. In Github, the \textit{Market} and \textit{Project} conventions -- somehow stronger in the non-AI texts -- are quite dominant, referring to licensing or commercialization for the first one and to the field of computer science, programming, and  software for the second one. 
In contrast with these two subsets, the \textit{Industrial} convention shows a lower percentage in Reddit, together with the \textit{Green} convention, while it is dominated by a cluster consisting of the \textit{Inspired}, \textit{Domestic},  \textit{Civic} (at least for the AI-texts), \textit{Project}, and \textit{Renown} conventions. Therefore, Reddit seems to be more balanced, due to the presence of a different set of conventions, reflecting the variety of topics and approaches in its discussions, while Github and S2 are dominated by one or two conventions. 
Generally speaking, the \textit{Green} convention is scarcely found (at least in Github and S2), showing that ecological and sustainable considerations are of little importance in these two subsets. The \textit{Market} convention often refers to questions of (commercial) licensing or business models, it was not excepted in the scientific articles, while it should be more present in software development.

The comparison of conventions of AI and non-AI samples reveals interesting tendencies for all three sources. By carefully looking at the results shown in figure \ref{fig:Github_convention_proportions}, a positive ratio can be observed between AI and non-AI domains for two conventions: the \textit{Domestic} and the \textit{Project} one. Only the \textit{Inspired} convention shows a negative ratio for all three subsets, confirming that AI related texts are more related to innovative and inspired approaches than non-AI ones. Interestingly, the ratio for the \textit{Industrial} convention differs between the three subsets with nearly no difference in Github, a positive ratio in Reddit and a negative one in S2, highlighting the importance of standardization and scientific methods

Figure \ref{fig:convention_cooccurences} shows the co-occurrences of conventions in the AI related items. The most interesting finding is the dominant correlation between the \textit{Industrial} and \textit{Inspired} conventions in the S2 subset, confirming its specific scientific character. In Reddit, validating the findings from figure \ref{fig:Github_convention_proportions}, we can observe a rather balanced proportion and co-existence of conventions, with slightly higher correlations in the combination of the \textit{Domestic} and \textit{Inspired} as well as the \textit{Domestic} and \textit{Project} conventions. This is in line with reflections on traditional and experienced-based ways of doing, as well as discussions on power and hierarchy, present in the Reddit subset. In contrast, Github shows a slight surplus in the combination of \textit{Industrial} and \textit{Inspired}, as well as \textit{Industrial} and \textit{Market} with percentages over $\sim$10\%, showing the content alignment of this subset. In none of the subsets, we find significant co-ocurrences with the \textit{Civic} convention, indicating a certain disconnection between civic values and the other dominant conventions in the AI domain.   
%
%

\paragraph{Qualitative sentence evaluation}

Automatic convention classification goes beyond merely detecting significant buzz-words. The correct attribution of a label has to include the buzz words, which refer to the ‘worth' of each convention. Additionally and more important it also must include a corresponding practical test (see \citet{boltanski2006}), which checks the corresponding ‘worth'. 
In the case of the \textit{Industrial} convention that is a procedural test, as any process can be only classified as \textit{Industrial} - in the sense of the EC - if it develops or produces something efficiently and productively in a standardized way.  
A label is only correct if this test is passed.

To illustrate this procedure and show the reliability of our classifiers on the basis of these requirements we compare a list of three sentences pairs (one pair per data source). The sentence pairs consist of one high accuracy (‘good example’) and one low accuracy (‘bad example’) sentence per data source 
from the \textit{Industrial} convention:

\begin{table}[h!]
    \centering\small
    \begin{tabularx}{\columnwidth}{lXr}
    
    & \textbf{\small{S2}} \\

    (1) &\textit{Graph partition can then be formulated as searching an optimal interface in the node weighted directed graph without user initialization.} &\faThumbsUp
     \\
    (2) &\textit{Effective soil mapping on farms can enhance yields reduce inputs and help protect the environment.} & 
    \faThumbsDown
    \\
    \midrule
    & \textbf{\small{GitHub}} \\
    (3) &\textit{It is often able to determine a good approximation of the true pareto front in significantly less iterations than genetic algorithms. } & 
    \faThumbsUp\\
    (4) &\textit{Full documentation is available at: docs.sypht.com repository is an apache licensed java reference client implementation for working with the api.started to get started you'll need some api credentials i.e a 'client-id' and 'client-secret'.} & 
    \faThumbsDown
    \\
    \midrule
    & \textbf{\small{Reddit}} \\
    (5) &\textit{They use it to model things like large scale particle interactions in a more computationally efficient way.} & 
    \faThumbsUp\\
    (6) &\textit{I would actually prefer if it generated Java code so I could tweak it by hand.} & 
    \faThumbsDown
    \end{tabularx}


\end{table}
In example (2)
from S2, the buzz-word ``effective'' does not automatically mean that this sentence belongs to the \textit{Industrial} convention. Simple technical descriptions such as example (4) from GitHub 
does also not imply any convention, although technical, scientific or industrial words are used. In contrast, (1) 
(extracted from S2) or 
(3)
(extracted from GitHub) include buzz-words, such as “approximation", “significantly"
, or “optimization" and they refer to standardized processes. Accordingly, they belong to the \textit{Industrial} convention. The Reddit example 
(5)
implies modelling as the central process for obtaining efficiency (corresponding with the industrial convention), while the example 
(6)
from the same data source does not refer to an industrial standardized process and therefore corresponds to the \textit{Domestic} convention.

We carry out several iterations of labelling, training and qualitatively analyzing the conventions. The analysis of sentences based on these conventions includes context information of the coded threads in order to determine the ‘practical test' and achieve a first step in grasping the social complexity of the EC in an automated classification.

\section{Discussion}

The EC and the automatic classification of the conventions offer a comprehensive insight into the dominant conventions and moral orders in the AI-field, partly linking and explaining the functioning of the five primary categories of motifs listed in \textit{Related Work \ref{sec:related_work}}. For instance the \textit{Inspired} convention can be associated with the categories of intrinsic motivation and learning (e.g. development of personal skills or knowledge), whereby the latter is also partly represented by the \textit{Domestic} convention. Furthermore, the category of external rewards can be attributed to the \textit{Market} convention and community recognition to the \textit{Renown} convention. An important finding in this regard is that the \textit{Industrial} convention, which turned out to be one of the most dominant ones in the subsets  investigated (see section "Evaluation of conventions"), is not reflected by any of these motifs.

There are ongoing discussions and research on the backgrounds and moral orders, which influence the development of the digital world. In this context, \citet{Castells2001} refers to the evolution of the internet as the result of the intersection of diverse cultures, from the purely ‘geek’ and technocratic to the outmost capitalist, melded with that of hackers and libertarians. The present study of the prevailing conventions in AI research, development and discussions continues and deepens this reflection, showing that there is a certain dominance of a techno-meritocratic culture (reflected in the \textit{Industrial} convention), at least in the scientific and technical descriptions of the AI projects. Less influence -- depending on the specific project and topic -- of the virtual communitarian culture (the \textit{Civic} and partly \textit{Project} and \textit{Green} conventions), the entrepreneurial culture (reflected in the \textit{Market} convention) and the hacker culture (the \textit{Domestic} and \textit{Inspired} culture).
In contrast, the Reddit subset includes blog posts, conversations and discussions on a variety of issues related to the field of AI, including ethical reflections, historical analysis, utopian and dystopian views. Hence, in the qualitative analysis \cite{Mayring2014} of the randomized sample of Reddit subset, pre-classified by the automatic classifiers and focusing on the concurrence of conventions (in the same sentence or in consecutive sentences), no dominance of one or two conventions is observable. Rather, Reddit seems to be characterized by a couple of specific co-occurring conventions, which seem to be central to the discussions around AI, indicating possible (ethical) conflicts. There seems to be, e.g., an ongoing conflict between the \textit{Industrial} and \textit{Domestic} convention around AI, reflecting discussions about the desirability and possibility to develop human-like machines or machine-like humans, and the superiority of human vs. AI. The EC and the automatic classifiers with its underlying concepts of standardization and optimization (in the case of the \textit{Industrial} convention) and trustworthiness, hierarchy and experience (in the case of the \textit{Domestic} convention) illustrates these conflicts. The automatic detection of conventions, as proposed by the classifiers, is able to shed light on the underlying moral assumptions in the AI (and other) fields. By this, it supports a deepened and mutual understanding of different points of view and moral backgrounds.

Our work involves a large amount of human knowledge and interaction. Accordingly, different types of bias might occur. \citet{Olteanu_2019} report a list of biases in areas such as to \textit{Data acquisition} and \textit{Data querying}, \textit{Data filtering} and also \textit{Biases in results interpretation} and \textit{Issues with the evaluation and interpretation of findings}. We briefly discuss the measures we take in this work to promote neutrality.
Due to the size of the content of both Github and Reddit, strong preselection is necessary. This is not the case for S2, where we gather the complete publicly available dataset and perform subsequent steps on the whole dataset.
We attempt to gather data from GitHub and Reddit in an equal manner. To ensure extensive discussion and good quality we collected data from repositories of all different levels of popularity (GitHub) and  all the threads with more than 4 upvotes (Reddit). 
%
%
To limit the bias in individual researchers' labeling in the active learning pipeline, the researcher triangulation and the sampling process from different levels of confidence both aim to mitigate this problem.
We evaluated the EC model with well-known performance metrics by convention and by data source to study potential systematic differences and incorporated qualitative analysis. 
We aim to foster reproducibility as well as discussion on methodological approaches so we release our dataset models and experiments to the research community.

\subsection{Limitations}
\todo[inline]{to discussion or limitations?}
We assume similar classifier performance on the AI and non-AI portions of the dataset although we do not carry out an empirical evaluation of non-AI portions of our dataset; the results for both the AI and non-AI portions in figure \ref{fig:convention_cooccurences} support this assumption. Furthermore, we assume the wording to be similar in the AI and non-AI portions of the dataset. Even as each data source belongs to a different text type, all data sources for both portions come from the computer science related technical domain. 
However, this assumption remains speculative and as such it would benefit from empirical evaluation on labeled sentences.

In the approach of this paper, items in the dataset are analyzed on sentence-level. According to the EC literature, conventions are better reflected on discussions where individuals need to defend their positions. Future work can focus in using current shape of the EC classifiers to analyze other data sources that, if having a conversational nature, will be better confronting and reflecting the conventions. 

Further, we have observed that the proposed techniques are highly dependent on the collection of high quality training data. Although an approach to facilitate such gathering has been proposed, further advances might be required to reduce the amount of manual work to be done by human annotators.

The EC is a social theory based on and therefore limited to Western political philosophy. Further, non-Western 'moral orders' are not reflected by the EC and the current analysis. But with further training of the models with non-Western-centric datasets, further conventions might be found, enriching not only the EC, but widening a global comprehension of morals.

\section{Conclusion and Future Work}

In this work, we described approaches both to analyze and predict conventions according to the EC. We created a dataset mainly from three text sources of scientific research: paper abstracts from scientific conferences and software development and analyzed the distribution of conventions in each sub-domain. We developed an interactive architecture based on active learning both to support domain experts in data labeling and select the most valuable items to train ML classifiers. Preliminary results on the ML classifiers trained on the EC showed promising results. In an additional study, the results were contrasted with the results from a classifier trained on software conventions and we have shown comparable and understandable results on both theoretic frameworks. 

The approach presented in this paper is the first contribution towards building an automatic text classifier of EC. The use of automatic models to perform the analysis enables the possibility of considering large amounts of information when accounting the conventions in a given dataset. This approach could be used in future analysis to extract conclusions in a variety of domains where prevalence of the EC needs to be studied. To facilitate the re-usage of this work, a repository
 containing the implemented code and the collected data has been published.

This work focused on three data sources which we considered relevant to reflect different perceptions about AI, i.e. the perspective of researchers, developers and the general public. In the future it would be interesting to study other types of interactions in data sources such as newspapers, online videos and chats.

In further steps, one focus will aim to detect and analyze common conflicts in software development and their underlying (assumable conflicting) conventions, beyond the already obvious problems of coordination between open source- and profit oriented AI development. With this, we hope to contribute to a more plural understanding of AI research and development, considering underlying moral registers which influence the motivations, objectives, processes and values of these projects. 

\bibliographystyle{aaai}
\bibliography{conflict_cooperation}

\end{document}